\documentclass{osa-article}

\journal{osajournal}

\articletype{Research Article}

\begin{document}

\title{Beating Optical Turbulence Limits Using High Peak-Power Lasers}

\author{Michael H. Helle\authormark{1,*}, Gregory DiComo\authormark{1}, Samantha Gregory\authormark{2}, Aliaksandr Mamonau\authormark{1}, Dmitri Kaganovich\authormark{1}, Richard Fischer\authormark{1}, John Palastro\authormark{3}, Scott Melis\authormark{4}, and Joseph Pe{\~n}ano\authormark{1}}

\address{\authormark{1}US Naval Research Laboratory, Washington DC
\\
\authormark{2}University of Alabama in Huntsville, AL
\\
\authormark{3}University of Rochester, Laboratory for Laser Energetics, Rochester, NY
\\
\authormark{4}Georgetown University, Washington DC

}

\email{\authormark{*}mike.helle@nrl.navy.mil} 



\begin{abstract}
We experimentally demonstrate the ability of a nonlinear self-channeling beam to resist turbulence-induced spreading and scintillation. Spatio-temporal data is presented for an 850-meter long, controlled turbulence range that can generate weak to strong turbulence on demand. At this range, the effects of atmospheric losses and dispersion are significant. Simulation results are also presented and show good agreement with experiment.
\end{abstract}

\section{Introduction}
Atmospheric turbulence is one of the dominant physical mechanisms limiting long distance optical propagation. Turbulence-induced inhomogeneities in the refractive index impart random phase front distortions that lead to spatial and temporal distortions in the optical field. When considering high-power laser pulses, the presence of turbulence leads to further complications due to the nonlinear nature of the propagation physics. For example, in strong turbulence turbulent fluctuations smaller than the size of the beam lead to the random formation of hot spots across the beam. These hot spots become self-reinforcing through nonlinear self-focusing and eventually lead to filamentation \cite{Braun:95, PhysRevE.66.046418, doi:10.1063/1.1648020, Diels:1034792}. Here whole-beam self-focusing cannot be achieved and the beam breaks up into a number of filaments approximately equal to the number of critical powers within the beam \cite{TALEBPOUR1999285, Fibich:04}. The generation and wander of plasma filaments in turbulence have been studied extensively \cite{Kandidov_1999, Chin2002, Ackermann:06, PhysRevA.78.033804}. Recently, it has been shown through simulations that this process reduces the control over the filamentation-onset distance. Turbulence reduces the mean onset distance while also increasing the statistical spread in the onset distance\cite{Penano:14}.

In this work, we present both experimental and simulation results for conditions where a high-power beam can self-channel though turbulence while resisting beam breakup. This work relies on the fact that realistic turbulence consists of eddies with a spectrum bound by an inner and outer scale size. The inner scale of turbulence is the result of the fluid viscosity preventing smaller scale eddies from forming. For beams much smaller than the smallest turbulent eddy, index-induced phase error tilt results in propagation where only beam-pointing fluctuations, known as wander, are present. For linear propagation over a long distance, natural diffraction increases the beam size to a point where this condition eventually can no longer be met. However, this is not the case for a nonlinear self-channeling beam, which can maintain its size over long distances by using nonlinear self-focusing to balance out diffraction \cite{PhysRevA.96.013829, Hafizi:17}. This condition is met when the peak power is equal to the critical power in air, $\sim$5GW\cite{Liu:05}.

\section{Experimental Setup}
High peak powers are typically accessed through short-pulse, and thus large-bandwidth, systems. The effects of group velocity dispersion in the atmosphere are critical to long distance propagation, and are a necessary consideration as part of this investigation. For instance, assuming a group velocity dispersion (GVD) coefficient of 22 fs$^{2}$/m at 800 nm\cite{Ciddor:96}, a 50 fs transform-limited pulse would double in length within the first 100 m of propagation in air. Such a propagation length is difficult to access in a typical laboratory environment; therefore, experiments were performed within the David Taylor Model Basin at NSWC Carderock, MD. This facility provided an indoor $\sim$1 km range with low ambient turbulence in which to perform controlled-turbulence propagation experiments. 

\begin{figure}[h!]
\centering\includegraphics[width=13cm]{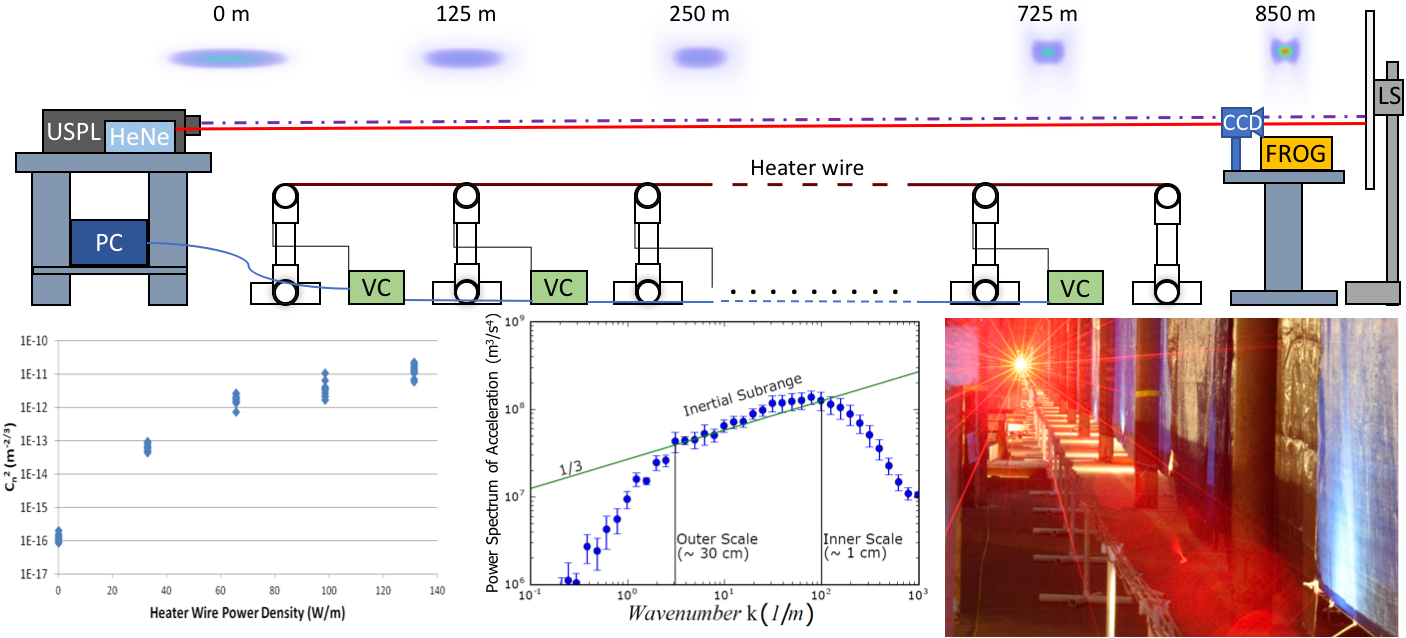}
\caption{Above: schematic of experimental setup with simulated intensity profiles showing compression of the negatively chirped pulse due to GVD. The turbulence generator is constructed of 12 separate zones controlled by individual voltage controllers (VC) that are networked to a PC. The ultrashort pulse (USPL) and HeNe laser beams co-propagate to the receiver end. The HeNe beam is captured by the LISSD diagnostic \cite{DiComo:16, CONSORTINI200319} for turbulence characterization. Meanwhile, the USPL is either imaged off the laser stop (LS) by the CCD for transverse beam profiling or captured by the second harmonic frequency resolved optical gating device (FROG) for temporal pulse measurements. Lower left: Path independent turbulence strength, C$_{n}^{2}$, as a function of heater-wire power density. Lower Center: Power spectrum of acceleration for turbulence produced at the beam height above the wires. The rolloff from the 1/3-power scaling indicates the inner and outer scales\cite{2018PhDT........35D}. Lower right: Photo of the propagation range with heater wires, laser curtains, and HeNe laser used for turbulence characterization.}
\label{Figure1}
\end{figure}

Phase plates are commonly used in laser propagation studies as a method to model the effects of turbulence in a controlled fashion. However, they are incompatible with high-peak power propagation due to the low optical damage thresholds and difficulties in scaling both the linear and nonlinear susceptibilities. Inspiration in how to address this problem instead came from how atmospheric turbulence is naturally produced: solar surface heating coupled with convective mixing. This process was emulated by using long electrical heating wires suspended below the laser path to heat the air and drive the mixing process. While this method is unable to produce a static phase-front distortion like a phase plate would, it does produce a controllable, well-defined statistical distribution of eddies that is consistent with those produced in nature. Initially presented in \cite{DiComo:16, Helle:17}, the design and characterization of this device, as installed for this experiment, are summarized in Figure \ref{Figure1}.

Owing to the long distances and large beam sizes used in this experiment, the turbulence generator consisted of twelve individual zones spanning a total of 825 m. Each zone consisted of four pairs of heater wires, evenly spread over a 10 cm width perpendicular to the beam propagation direction. Each zone is independently computer controlled, with thermocouples to monitor individual zone temperatures. For the present experiments, each zone was operated at the same setpoint; however, they could be individually controlled to produce various longitudinal turbulence profiles. The generator was suspended under tension 60 cm from the ground by PVC supports, with the laser beam placed an additional 70 cm above the wires. This provided sufficient distance for the turbulence to fully develop into a von Karman spectrum ($\sim$1 cm inner and $\sim$30 cm outer scale, see Figure \ref{Figure1}) and provide a transverse zone over which the beam would encounter uniform statistics. The LISSD diagnostic discussed previously \cite{DiComo:16,CONSORTINI200319} was used to measure the path-averaged turbulence strength and inner scale. For this work at the Carderock facility, the baseline scintillation index was 0.23 and was varied up to 7.4. Here we define scintillation index as the variance of the intensity normalized to the square of the mean.

A Coherent Astrella 1k-USP Ti:Sapphire laser (7 mJ, 35 fs, 800 nm, 1 kHz) was placed 12.5 m from the start of the generator. The output of the system is nearly collimated with a beam radius of 7 mm (measured at the 1/e$^{2}$ of peak intensity.) This is smaller than the measured inner scale and provides $\sim$6 Rayleigh lengths over the entire propagation length for examining self-channeling of the beam. A negative chirp is applied to the pulse in order reduce the peak power, while also using group velocity dispersion to compress the pulse during propagation. The optimal pulse duration used for these experiments is measured as 2.4 ps FWHM (P$_{0}$ = 2.3 GW, equal to P$_{Crit}$ for long pulses \cite{PhysRevA.85.043820}). This was necessary to maintain the self-channeling condition despite diffractive, absorptive, and scattering losses. Absorption and scattering accounted for a 12$\%$ loss of laser energy over the propagation range. 

The beam is characterized at the receiver end, 12.5 m beyond the end of the turbulence generator, at a total propagation distance of 850m. The pulse`s transverse and temporal profiles were characterized both at the transmitter and receiver using CCD cameras and a second harmonic frequency resolved optical gating (FROG). The FROG was used to measure temporal evolution due to GVD as well as nonlinear propagation effects such as self-phase modulation\cite{Boyd:640132}. 

\begin{figure}[h!]
\centering\includegraphics[width=13cm]{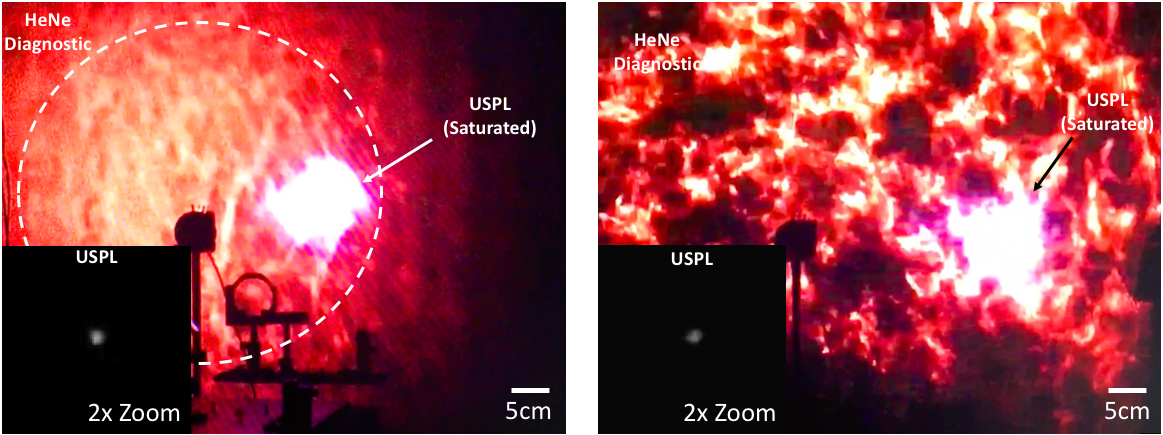}
\caption{True color images taken at 850 m showing the effects of turbulence on the beams with heater wires off, left, and at full power, right. The HeNe beam used to characterize the turbulence appears in red while the intense USPL beam appears white due to saturation of the camera. Representative, unsaturated images of the self-channeled beam from the receiver plane CCD are included as insets. Note the significant scintillation and increase in beam size of the HeNe while the USPL is comparatively unaffected.}
\label{Figure2}
\end{figure}

\section{Results and Discussion}
Receiver plane images at low (scintillation index = 0.23) and high (scintillation index = 7.4) turbulence are shown in Figure \ref{Figure2}. The typical effects of turbulence, namely intensity fluctuations (scintillation) and beam spreading, are clearly present in the linearly propagating HeNe, while less obvious in the self-channeled beam. Quantitative results for the pulse properties follow.

To compare to measurements with nonlinear propagation predictions, we use the HELCAP code \cite{PhysRevE.66.046418}, which solves the paraxial equation for the laser electric field envelope A:

\begin{equation}
    [2 i k (\frac{\partial}{\partial z} + \alpha) - k \beta_{2} \frac{\partial^{2}}{\partial \tau^{2}} + \nabla^{2}_{\perp}] A = -2 k^{2}_{0} \delta n A,
\end{equation}

where $\delta n$ = $\delta n_{T}$ + $n_{s} I$, $z$ is the axial propagation coordinate, $\tau = t - z/vg$, $vg$ is the pulse group velocity, $\alpha$ is the atmospheric extinction coefficient (e.g., scattering and absorption), $\beta_{2}$ = 0.22 fs$^{2}$/cm is the group velocity dispersion (GVD) coefficient at $\lambda$ = 800 nm, $n_{2}$ = 3.9x10$^{-19}$ cm$^{2}$/W is the long-pulse Kerr index of air, and $\delta n_{T}$ is the refractive index due to turbulence, which we model as a series of random phase screens with Tatarskii spectral characteristics \cite{1971etaw.book.....T}. We solve the above equation using a split-step method in which linear terms are handled in Fourier space and nonlinear terms are handled in coordinate space \cite{10.1007/3-540-46629-0_9}. We use a Cartesian grid with 512x512x512, 400 steps in the propagation coordinate z, and 40 uniformly spaced turbulence phase screens along the beam path. Ensemble-averaged quantities are obtained using on the order of 10$^{3}$ difference realizations of turbulence. To obtain the C$_{n}^{2}$ values used in the simulations, we used analytic equations for the scintillation index of a Gaussian beam \cite{Andrews_2005} and applied them to the propagation of the HeNe probe beam used in the experiments. Using C$_{n}^{2}$ as a free parameter, we fit the analytic expressions to the experimentally measured scintillation indices. This resulted in C$_{n}^{2}$ values from 2.3x10$^{-14}$ to 1.5x10$^{-12}$ m$^{-2/3}$. The inner scale was taken to be 2 cm, which is comparable to the experimentally measured value, and outer scale was taken to be 1.5 m, approximately equal to the beam height above the ground. 

We simulated the propagation using a 6.5 mJ beam, with the experimentally measured transverse beam intensity profile and longitudinal field profile as initial conditions in the simulation. To account for possible wavefront distortions in the initial pulse, we introduced a 30 $\mu$rad defocusing contribution such that the simulated and experimentally measured spot size at the receiver end agreed in conditions of lowest turbulence. In addition, we assumed a pointing jitter of 8 $\mu$rad, which is the manufacturer`s specification for the laser system.

\begin{figure}[h!]
\centering\includegraphics[width=9cm]{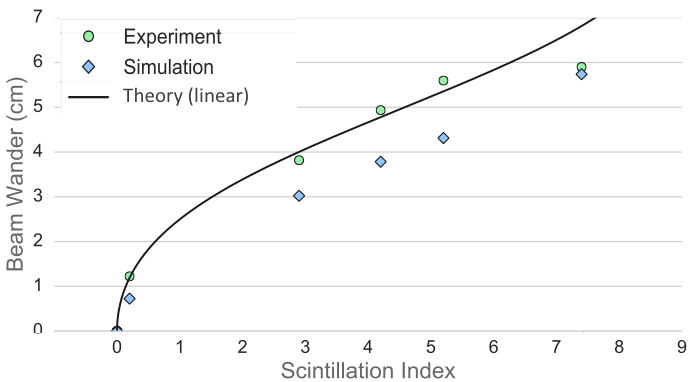}
\caption{Beam wander as a function of turbulence strength. Results from experiments, simulation and turbulence theory for linear propagation (solid line) are plotted.}
\label{Figure3}
\end{figure}

As stated earlier, a beam smaller than the inner scale of turbulence will resist spreading and breakup, but will still experience turbulence-induced beam wander. This is clearly seen in Figure \ref{Figure3}, where beam wander extracted from experiments and simulations are is compared to linear turbulence theory\cite{Andrews_2005}. All three agree well with the overall trend that as turbulence strength is increased, beam wander also increases. 

\begin{figure}[h!]
\centering\includegraphics[width=10cm]{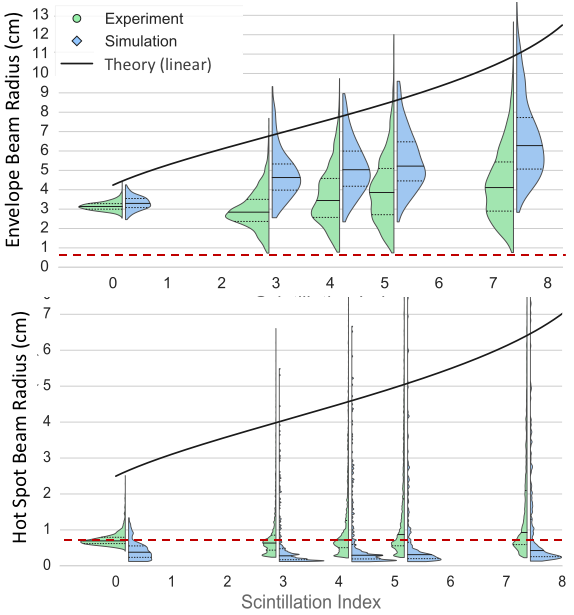}
\caption{Probability distributions of beam envelope rms radius, top, and hot spot rms radius, bottom, as a function of turbulence strength. Within the distributions, the mean is indicated with a solid line, while the one standard deviation points are marked with dashed lines. The corresponding theoretical mean spot size for purely linear propagation is plotted as a solid line. Additionally, a dashed red line is included to show the initial beam size measured at the transmitter.}
\label{Figure4}
\end{figure}

For the self-channeling condition to be met the beam must remain smaller than the inner scale of turbulence or the transverse coherence length over the entire propagation range \cite{PhysRevA.96.013829}. Probability distributions of the beam size for varying turbulence strength are plotted in Figure \ref{Figure4}. As with our previous work \cite{2018PhDT........35D, doi:10.1063/1.4821447}, plots of both beam envelope and hot spot are included to highlight the self-channeling property. When examining the beam envelope, we observe that even in weak turbulence the beam envelope is larger than the initial beam size. This is due to the temporal profile of the pulse, where there exist regions below the critical power in the front and back of the pulse. Energy within these regions tends to undergo natural diffraction leading to a beam envelope larger than the self-channeled beam size. Even with this effect, the beam envelope is significantly more resistant to beam spreading than linear theory predicts. At the highest turbulence strength, the mean envelope size increases by 33$\%$, compared to 225$\%$ predicted by theory. Simulation results are in good agreement with the experiments. The simulations tend to predict a larger mean envelope size except at the lowest value of turbulence. In all cases, experiment and simulation show a very similar width at one standard deviation.

In addition to the envelope statistics, we applied a 50$\%$ threshold to the beam images to extract the behavior of the beam`s hot spot. We observe that the peaks of the distributions line up with the initial spot size at all levels of turbulence strength. More importantly, the mean hot spot size matches the initial spot size, except at the highest levels of turbulence where it is still within one standard deviation. The simulation results are in general agreement, but tend to indicate additional self-focusing of the beam. This is likely the result of a non-ideal initial phase front not captured in the simulations. 

\begin{figure}[h!]
\centering\includegraphics[width=10cm]{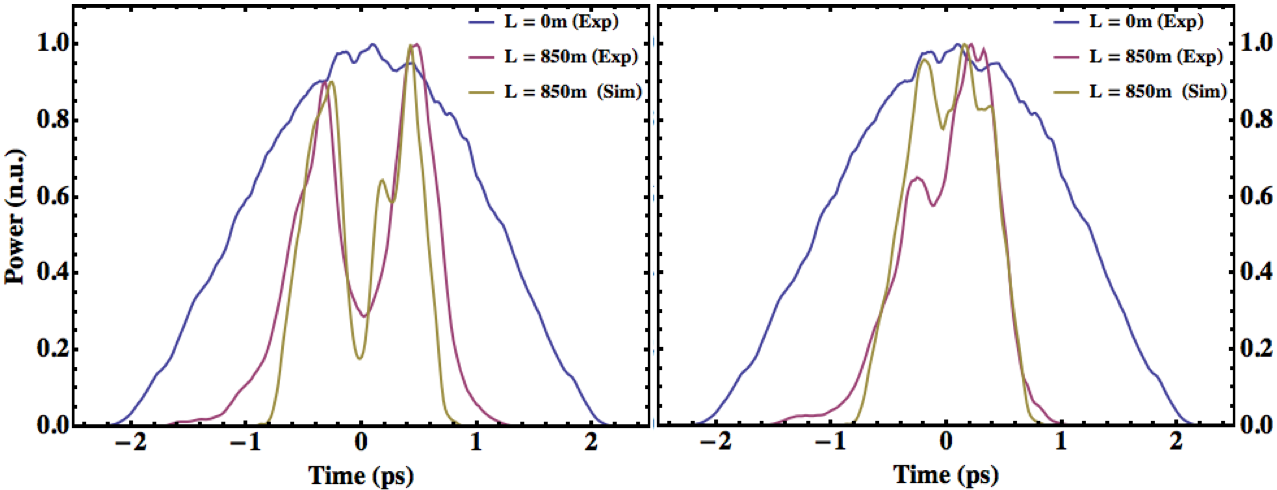}
\caption{Left: Temporal profile of the self-channeled beam for the case when linear compression dominates and pulse splitting is not observed. Experimental curves for the initial profile and profile at the end of the propagation range are included and compared to simulations. Right: Temporal profile for the case when pulse splitting is observed.}
\label{Figure5}
\end{figure}

The discussion to this point has been limited to transverse beam properties; however, the broadband nature of the pulse requires an examination of the temporal dynamics. As pointed out in \cite{PhysRevA.96.013829}, temporal compression is necessary to maintain the self-channeling condition when absorption, scattering, and diffractive losses are present. We observed instances in which the beam undergoes almost purely linear compression of the negatively chirped pulse due to GVD, as well as instances where pulse splitting is present, a clear signature of self-phase modulation. Temporal field profiles extracted from the FROG are included in Figure \ref{Figure5} with representative simulation results. The pulse compresses by a factor of $\sim$2.5, consistent with the compression expected using GVD values from the literature\cite{Ciddor:96}. Unfortunately, due to beam wander about the entrance aperture of the FROG, a rigorous statistical comparison cannot be made between experiment and simulation. Simulations indicate that temporal splitting is observed for turbulence realizations that yield smaller spot sizes (and thus greater intensities) at range; this is consistent with established physics of self-phase modulation. 

\begin{figure}[h!]
\centering\includegraphics[width=10cm]{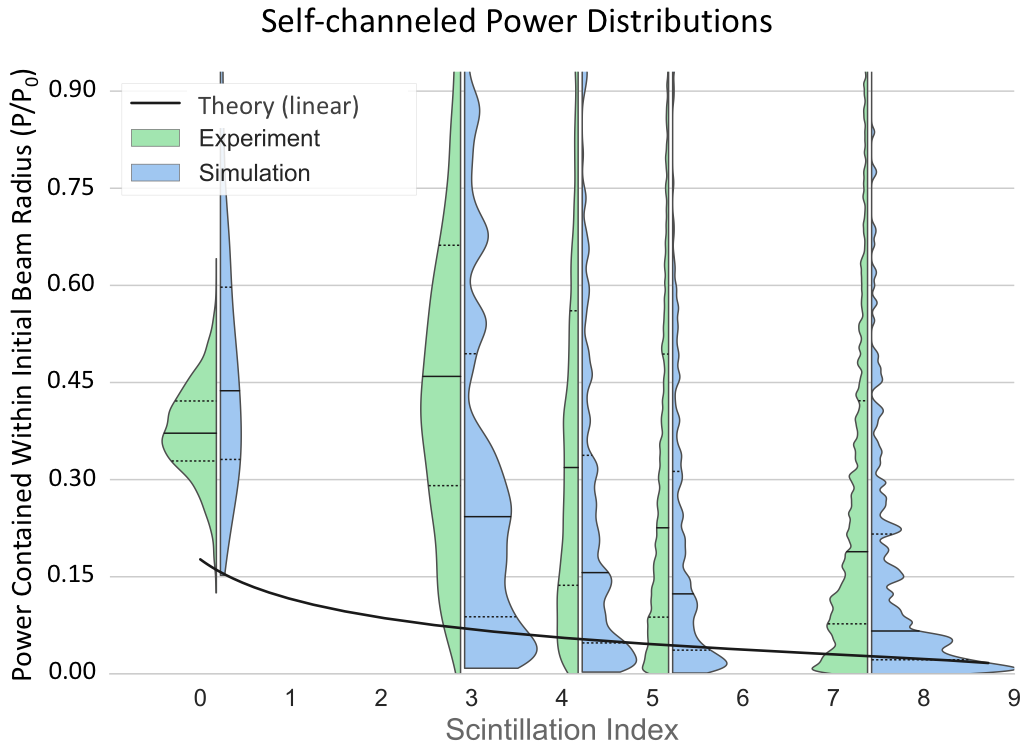}
\caption{Probability distributions of peak power contained within initial beam spot size. The distributions are calculated by taking the energy contained within the initial RMS beam radius divided by the measured pulse duration and normalized to the initial power, P$_{0}$. For the USPL, P$_{0}$ is equal to P$_{Crit}$ in air. Here 0.86 denotes a Gaussian beam containing P = P$_{0}$.}
\label{Figure6}
\end{figure}

The self-channeling is further highlighted by examining the channeled beam power, or the amount of energy contained within an area equal to the initial beam size centered on the beam centroid divided by the pulse duration, shown in Figure \ref{Figure6}. Examining the linear theory curve, the combined effects of diffraction, GVD, and turbulence-induced spreading are expected to lead to a beam power of <0.15 P$_{0}$. By comparison, the self-channeled beam produces a mean fractional energy upwards of 0.45 P$_{0}$ at moderate turbulence strength, and 0.2 P$_{0}$ at the highest turbulence strength; an $\sim$7x increase compared to a purely linearly propagating beam. Simulations are in general agreement, but tend to underestimate self-channeled power. The reason for this is the simulations tend to predict an overall larger beam, see Figure \ref{Figure3}.

In this work, we have shown that a nonlinear self-channeled beam is capable of resisting turbulence-induced beam spreading. This is permitted for beams where the beam size is smaller than the inner scale size or transverse coherence length. Both experiments and simulations confirm that while such a beam obeys theoretically predicted beam wander similar to that of a linear beam, the mean RMS spot size can be maintained to within 33$\%$ of the original spot size even in deep turbulence conditions with scintillation index >7. Additionally, the hot spot beam size is maintained to within 18$\%$ of its original value over the entire range of turbulence strengths examined.  Finally, it has been shown that the self-channeling behavior can be maintained in realistic conditions where the effects of dispersion and losses are present with >20$\%$ of the initial peak power self-channeled even in deep turbulence; an increase of >7 times that of purely linear propagation. These results are fundamentally exciting in that self-channeling beams provide a means to significantly reduce beam spreading which limits optical beam transmission in random media. 

We would like to thank Andreas Schmitt$-$Sody, Jennifer Elle, Daniel Gordon and Bahman Hafizi for useful discussions, Thomas Buetner for suggesting the Carderock Facility, and especially the entire David Taylor Model Basin staff for their tireless help. This work was made possible through funding by the Joint Transition Office $-$ High Energy Laser, the Office of Naval Research, and Naval Research Laboratory Base fund.


\bibliography{HelleNonlinearChanneling}






\end{document}